 \documentclass[authorversion]{acmart}

\newcommand{\tldr}[1]{\textcolor{magenta}{[TLDR: #1]}}

\newcommand{\SP}{\textbf{\textcolor{red}{[SP]}}}

\usepackage{multirow,array}

\usepackage{mathdots}
\usepackage{amsthm,amsmath}
\usepackage{mathtools}
\usepackage[ruled, noend]{algorithm2e}
\usepackage{cleveref}
\usepackage{graphicx}
\usepackage{lscape}

\crefname{algorithm}{algorithm}{algorithms}
\Crefname{algorithm}{Algorithm}{Algorithms}

\crefname{line}{algorithm}{algorithms}
\Crefname{line}{Algorithm}{Algorithms}

\AtBeginDocument{%
  \providecommand\BibTeX{{%
    \normalfont B\kern-0.5em{\scshape i\kern-0.25em b}\kern-0.8em\TeX}}}

\setcopyright{acmcopyright}
\copyrightyear{2025}
\acmYear{2025}
\acmDOI{XXXXXXX.XXXXXXX}

\acmPrice{15.00}
\acmISBN{978-1-4503-XXXX-X/18/06}

\begin{document}


\title{Towards a Better Modqueue: Designing for Diversity Across Moderator Objectives and Workflows}



\author{Tanvi Bajpai}
\email{tbajpai2@illinois.edu}
\affiliation{%
  \institution{University of Illinois Urbana-Champaign}
  \city{Urbana}
  \state{Illinois}
  \country{USA}
  \postcode{61801}
}

\author{Eshwar Chandrasekharan}
\email{eshwar@illinois.edu}
\affiliation{%
  \institution{University of Illinois Urbana-Champaign}
  \city{Urbana}
  \state{Illinois}
  \country{USA}
  \postcode{61801}
}

\renewcommand{\shortauthors}{T. Bajpai and E. Chandrasekharan}

\begin{abstract}


Reddit relies on volunteer moderators to enforce community rules, configure tools, and review flagged content. This labor is substantial, worth millions in unpaid effort, and increasingly hard to sustain as communities grow. While recent updates to Reddit’s modqueue emphasize efficiency and reducing redundancy, recent research shows that moderators use the interface in varied ways, value objectives beyond throughput (such as fairness and accuracy), and often resist features that disrupt workflows. In this paper, we survey 106 active Reddit moderators to examine the objectives they bring to their modqueue work and the kinds of interventions they consider helpful. Our findings highlight wide variation in values and workflows, with no single objective beyond accuracy dominating, and different perspectives on which interventions are useful. To address this diversity, we introduce a simulation-based approach that can complement empirical findings by probing tradeoffs and testing potential interventions, and provide design recommendations based on our findings.



\end{abstract}






\maketitle


\section{Introduction} 

Reddit utilizes a community-reliant model of moderation, in which there is a heavy reliance on volunteer community moderators (\emph{mods}). These volunteer mods are responsible for establishing community-specific rules and guidelines~\cite{reddy2023evolution}, configuring automated moderation tools~\cite{jhaver19automod}, and making case-by-case judgment calls regarding whether to remove potentially problematic or rule-violating content posted in their communities \cite{matias2019civic, chandrasekharan19crossmod}. While this model allows individual communities the flexibility to enforce and uphold their own unique norms~\cite{chandrasekharan2018norms} (especially compared to more centralized approaches like those used on Instagram or TikTok) it also means that volunteer mods shoulder a significant share of the responsibility for maintaining sometimes large and constantly growing communities~\cite{kiene2019volunteer}---that too, without compensation~\cite{li2022all}. However, the effort expected from volunteers to maintain communities is showing signs of becoming unsustainable: \citet{li22monetary} estimate the monetary value of the work done by Reddit’s volunteer mods in 2020 to be over 3 million USD, raising questions about whether moderation can reasonably be expected to remain volunteer labor. At the same time, \citet{park2022measuring} find that most antisocial content on Reddit remains unmoderated, suggesting potential shortcomings across different parts of the community-reliant moderation process. In particular, such gaps could arise from underreporting by community members, ineffective flagging tools, or bottlenecks in moderator review, i.e., when rule-violating content is correctly flagged but awaits moderator judgment before removal.

While prior work has examined issues with user reporting \cite{seering2022metaphors, gilbert2023towards, koshy2023alignment} and explored how automated flagging tools might be improved with AI \cite{chandrasekharan19crossmod,zhan2025slm}, relatively little attention has been paid to improving the process of \emph{moderator review}. By \emph{moderator review}, we refer to the process where mods assess reported or flagged content and decide whether it is rule-abiding (and can be or remain posted) or rule-violating (and should be removed). This gap is notable because moderator review is one of the main mechanisms through which community norms and human judgment are applied in moderation decisions~\cite{jhaver19automod, chandrasekharan19crossmod}, and is therefore integral to community-reliant moderation models. Thus, there is a critical need not only to improve moderator review for more effective outcomes (e.g., ensuring that antisocial behavior does not go undetected) but also to ensure that mods are meaningfully supported in carrying out this work, so that volunteer labor becomes more manageable and the human-centered nature of community moderation is preserved.

\subsection{Defining What It Means to Improve Moderator Review on Reddit}

 On Reddit, a primary interface for moderator review is the moderation queue (or \emph{modqueue}), a consolidated listing of reported or filtered items awaiting moderator review \cite{modqueue}. 
 As such, one way to improve moderator review on Reddit could be to improve the modqueue interface itself through design changes that better support mods' review workflows. 
 
 However, what it means to actually ``improve'' the modqueue or ``better support'' mods' workflows remains unclear. A recent study found modqueue practices and workflows vary widely depending on factors such as community attributes, tooling preferences, and underlying moderation philosophies, among others \cite{bajpai2025queue}. Findings from the study also suggest that new features that have already been introduced by Reddit to improve review work in the modqueue  (aimed at improving efficiency and reducing redundancy \cite{reddit2024newmodqueue}) appear to be underutilized. The study asserts that this may be because these new features fail to align with mods’ established workflows or preferred tools~\cite{bajpai2025queue}. Another possibility is that efficiency and redundancy are not the objectives mods themselves view as most important. 
 
In this paper we focus on clarifying what ``improvement'' means in the context of the modqueue. Motivated by the aforementioned findings and observations, we approach this by examining two key ingredients that appear to shape how improvements should be understood and developed: (1) what objectives matter to mods during their review work, and (2) whether potential features or interventions align with, or disrupt, moderators' existing workflows.

 \subsection{Summary of Contributions}

In this paper, we begin by investigating what ``improvement'' to the modqueue means to moderators through a survey of 106 Reddit mods. The survey asks about the objectives they value during modqueue review, and their perceptions of potential design ideas and interventions. These findings establish the first two contributions of this work:

 \begin{enumerate}
    \item \textbf{Identifying objectives to design for.} Drawing from prior moderation research, we surface a set of objectives that mods value in the broader context of moderation. We then ask mods to evaluate the importance of each, provide their perspectives on how they matter for modqueue work, and describe any additional objectives they consider relevant.
     \item \textbf{Assessing intervention preferences.} Building on findings from prior work~\cite{bajpai2025queue} which indicated that collisions remain a persistent issue (despite new features designed to help mods avoid them), we propose a set of alternative intervention ideas aimed at helping mods to circumvent common issues. We then ask mods for their views on these ideas, probing whether such features would fit into their existing workflows and whether they would be perceived as useful. We also examine intervention modalities that could be implemented consistently across Reddit’s different interfaces (i.e., Old Reddit, New Reddit, and the mobile app \cite{modqueue,bajpai2025queue}), and highlight the kinds of information and design features mods would find most helpful to have surfaced within the modqueue.  
    
 \end{enumerate}
 
Our survey findings ultimately surface several tensions that complicate how ``improvement'' can be defined and designed for in the modqueue. This motivates the third and final contribution of our paper: 

\begin{enumerate}
\setcounter{enumi}{2}

    \item \textbf{Developing a simulation-based audit tool.} We introduce a modular agent-based simulation tool that enables experimentation with factors
    that are otherwise difficult to study directly and can shape the modqueue review process.
    We demonstrate how the tool can be used to audit current modqueue practices, provide insight into the efficacy of new modqueue interventions, and surface the trade-offs that may emerge as a consequence of adopting them.
\end{enumerate}

Findings from our survey and simulation experiments have implications for improving the design of moderator review systems like Reddit's modqueue.
We also discuss how HCI researchers studying complex processes like content moderation can benefit from agent-based simulations to audit and explore intervention strategies.

\section{Background}\label{sec:back}

\subsection{Online Community Moderation}

\citet{grimmelmann2015virtues} defines moderation of online communities as ``the governance mechanisms that structure participation in a community to facilitate cooperation and prevent abuse.'' While many platforms facilitate this governance by enforcing uniform, platform-wide policies or deploying algorithmic content filters \cite{gillespie2018custodians}, others rely more directly on the efforts of their members. Platforms such as Reddit, Discord, and Twitch adopt \emph{community-reliant} models of moderation \cite{caplan2018content,cai2023understanding}, where rules and guidelines are set locally and volunteer moderators are responsible for implementing and enforcing them. This approach ensures that communities can interpret norms in ways that reflect their own goals and contexts, but it also introduces significant variation across communities in how moderation is carried out \cite{fiesler2018reddit,chandrasekharan2018norms,weld19values,gilbert20historians}.

Beyond community norms and context, mods' personal practices and philosophies can also shape how moderation is carried out.  also These perspectives shape the kinds of actions moderators take in practice. Some moderation work is punitive, such as removing harmful content or sanctioning repeat offenders, while other work is constructive, aimed at encouraging pro-social contributions and reinforcing positive behavior \cite{bruckman1994deviant,chandrasekharan2022quarantined,seering2017pro,seering2019engagement,lambert24positive}. Which kinds of actions are applied, and in what context, often depend on community priorities and moderator philosophies, producing a wide range of strategies even when moderators share similar overarching goals.

Volunteer mods take on a variety of roles and responsibilities to moderate their communities \cite{seering2019engagement, gillespie2018custodians,matias2019civic}. One of the primary responsibilities of mods is to mitigate rule-violating and/or abusive behavior within the community \cite{bruckman1994deviant, gillespie2018custodians}. In a study conducted on moderation practices in Twitch chatrooms, \citet{seering2017pro} identify the ways in which mods encourage pro-social behavior and discourage anti-social behavior. One such method of deterrence was issuing reactive bans to users that engage in problematic behavior. \citet{kiene2016newcomers} found that removing rule-violating content quickly and consistently was integral for aiding mods in establishing norms for newcomers. In addition, several studies have explored how mods come to develop and establish community rules and norms~\cite{zhang2020policykit,fiesler2018reddit}. 

Researchers have also studied how moderation teams on platforms like Reddit often use automated tools and interventions to help remove problematic content \cite{seering2019engagement,jhaver19automod,chandrasekharan19crossmod}. Reddit's AutoModerator (or Automod) is a bot that can be configured by mods to automatically flag and/or remove certain types of content \cite{seering2019engagement, jhaver19automod}. \citet{chandrasekharan19crossmod} developed Crossmod, an AI-driven tool similar to Automod that utilizes cross-community learning, and newer work by \citet{zhan2025slm} explores how LLM's can be used to improve moderation tools. Despite the use of automated approaches to triage norm violating content for review, massive amounts of norm-violating content on Reddit remains unmoderated: \citet{park2022measuring} found that  only one in ten comments that exhibit anti-social behavior are actually getting removed by Reddit mods. This may because of issues surrounding user reports: \citet{10.5555/3632186.3632213} found that users sometime hesitate to report rule-violating content if doing so infringes on their privacy. Other studies show that some reports are made in bad faith \cite{gilbert2023towards,are2025flagging}, casting doubt on their reliability as signals that mods can rely on to address rule-violating behavior.

\subsection{Moderator Review on Reddit and the Modqueue}\label{sec:back:modq}

Moderator review on Reddit occurs when volunteer moderators evaluate flagged content and decide whether it aligns with their community’s rules and norms. Items requiring review reach moderators in several ways: through user reports, automated filters configured with AutoModerator, or platform-wide spam detection systems \cite{modqueue}. This work can be done directly from posts or comment threads, but Reddit provides the \emph{modqueue} as its central interface for aggregating all flagged items across a moderator’s communities. From the modqueue, moderators can take a range of actions, including approving or removing content, marking items as spam, locking threads, or applying flairs. Because it consolidates different streams of reports, the modqueue functions as a primary hub of moderator review on Reddit.

Notably, there is no single ``canonical'' modqueue: Reddit supports three interface versions (Old Reddit, Reddit.com or \emph{New Reddit}, and mobile), each of which presents reports differently and surfaces different contextual signals \cite{modqueue,pardes2018inside,reddit2024newmodqueue}. New Reddit, for example, introduces contextual panels and activity indicators intended to streamline review processes, while Old Reddit retains a simpler list-based design. Many moderators supplement these official tools with browser extensions such as Mod Toolbox \cite{modtoolbox2024} or Reddit Enhancement Suite \cite{res2024}, as well as custom bots built on the Reddit API. As a result, moderator review often involves navigating across multiple toolchains, with moderators tailoring combinations of features to their own habits, devices, and community needs \cite{bajpai2025queue}.

Recent work \cite{bajpai2025queue} investigating how Reddit moderators utilize the modqueue found that practices vary widely. Some moderators treat the queue as a checklist to clear, while others use it strategically to detect broader patterns in community activity. Mods also have different ways of selecting items from the queue to review first: many move sequentially through it, but others report prioritizing by report age, source (user reports vs. Automoderator), or content type (posts vs. comments). Even with added contextual panels, most moderators still leave the modqueue to gather surrounding conversation, review user histories, or consult teammates. Persistent challenges include coordination breakdowns, collisions when multiple moderators act on the same item, limitations in sorting and filtering features, and reliance on fragmented tooling preferences that can span multiple Reddit interfaces and third-party extensions.

\subsection{Moderation Objectives}\label{sec:back:objs}

Since one of the central aims of this paper is to examine which kinds of objectives moderators bring into their work with the modqueue, and to clarify what it means to “improve” modqueue practices, we present a brief review of objectives that have surfaced in prior moderation research. Across the literature, researchers and practitioners highlight a recurring set of concerns that provide useful grounding for our study.

A longstanding theme in moderation research is the pressures of efficiency. Studies consistently document the sheer volume of labor involved, whether in paid professional settings \cite{gillespie2018custodians} or in volunteer-based ecosystems such as Reddit \cite{matias2019civic,li22monetary,seering2019engagement,gilbert20historians}. The scale of this effort has motivated interventions that aim not just to increase speed but also to address urgency, ensuring that time-sensitive issues are resolved before they escalate \cite{kiene2016newcomers,chandrasekharan2022quarantined}. More recent investigations of modqueue use reinforce these themes: moderators describe report backlogs as stressful, and collisions (two mods handling the same report simultaneously) as frustrating wastes of limited capacity \cite{bajpai2025queue}. Together, these studies highlight efficiency and workload management as important for sustaining moderation over time. Similarly, efficiency and reducing redundancy in review work were objectives mentioned in new Reddit modqueue feature announcements and recent studies regarding specifically modqueue work \cite{reddit2024newmodqueue,bajpai2025queue}.

Some moderation research emphasizes the importance of maintaining moderator well-being. Content moderation demands extensive emotional labor \cite{matias2019civic,wohn19emotional,dosono2019reddit} and can involve exposure to harmful or distressing content \cite{steiger21wellbeing,Chen_2014}. Research has documented burnout, particularly in marginalized communities where moderators shoulder disproportionate burdens \cite{dosono2019reddit}, and surveyed interventions aimed at reducing harm and supporting wellness \cite{steiger21wellbeing}. 

Accuracy and fairness are also referenced as objectives that are important: Prior studies stress that rule enforcement accurately reflect community rules, since inaccurate or unexplained removals undermine trust and hinder norm establishment \cite{kiene2016newcomers,jhaver2019did}. Users frequently frame disputed removals as unfair, particularly when the rationale is opaque \cite{jhaver2019did}. Finally, several studies highlight the importance of agreement or consensus in decision-making. Research on online juries and expert panels suggests that group deliberation can increase legitimacy and consistency compared to individual or automated judgments \cite{hu2021can,pan2022comparing,gordon2022jurylearning,koshy2024venire}. Within volunteer teams, agreement helps ensure that enforcement decisions are not only accurate but also seen as fair to both members and moderators themselves \cite{lampe2004slashdot}.

\section{Survey Methods}

We designed and implemented a survey study whose findings make up the first two contributions of this paper: identifying moderators’ objectives for modqueue work and assessing the viability of potential interventions. 
This study was reviewed and approved by the Institutional Review Board (IRB) at the first author’s institution. Participants provided informed consent before beginning the survey, and incomplete responses were excluded to allow participants to withdraw consent partway through if they so desired.

\subsection{Survey Design}

We designed our survey to accommodate a wide range of moderator perspectives and experiences, ensuring that questions were broad enough to be relevant across different experience levels, community contexts, and modqueue interface preferences and practices. It consisted of multiple-choice and Likert-scale questions with strategically placed open-ended prompts.

The survey began with background questions about respondents’ moderation experience and the tools they use, including the degree to which they utilize the modqueue interface. Those who reported that they do not use the modqueue in any aspect of their moderation workflows were directed to the end of the survey. Those who reported using the modqueue were shown all of the closed-ended items in our survey, with conditional open-ended questions displayed based on their responses. Our survey questions each fell in to one of two categories that correspond to the first two contributions of this paper: Objectives and Interventions. We detail the specifics of the questions asked in each of these categories below. 

\subsubsection{Objectives}
Drawing on prior work in online moderation (\Cref{sec:back:objs}), we selected a set of seven objectives that could potentially be important to moderators in the context of mod review. We presented each objective as a short prompt describing the objective in general but easy-to-understand terms (see \Cref{tab:mod-obj-prompts}), and asked them to specify the degree to which they found each statement to be important. Respondents were asked to rate how important each was during modqueue review work using a five-point Likert scale ranging from ``Not at all important'' to ``Extremely important.'' When a moderator selected either endpoint, the survey displayed a conditional open-ended follow-up asking them to explain why they selected their importance rating. We also included a final open-ended item inviting respondents to provide additional objectives they considered important but that were not listed.

\begin{table}[h]
\centering
\begin{tabular}{|l|p{10.5cm}|}
\hline
\textbf{Objective} & \textbf{Prompt Shown to Participants} \\
\hline
Efficiency & Minimize the total time it takes to clear all active reports. \\
\hline
Accuracy & Ensure that reports are addressed accurately. \\
\hline
Toxicity Exposure & Limit continuous exposure to toxic or severe reported content. \\
\hline
Fairness & Ensure moderation decisions are perceived as fair by moderators and the community. \\
\hline
Consensus & Foster agreement among the majority of the moderation team on decisions. \\
\hline
Wellbeing & Ensure moderators’ wellbeing is not harmed in the moderation process. \\
\hline
Redundancy & Avoid instances where multiple moderators are working on the same report simultaneously \\
\hline
\end{tabular}
\caption{Objectives and corresponding prompts shown in the survey. We note that the objective names (e.g. ``Efficiency'' or ``Consensus'') were not shown to moderators to avoid confusion.}
\label{tab:mod-obj-prompts}
\end{table}

\subsubsection{Proposed Interventions}
To obtain moderator insights into potential ways of designing interventions, we began by asking moderators to reflect on features that could support objectives Reddit itself had highlighted in recent updates to the modqueue: specifically, improving efficiency and reducing redundancy \cite{reddit2024newmodqueue}. To ensure the questions were relevant across all interface versions (since newer Reddit features may not exist in the Old Reddit modqueue interface or the mobile app \cite{bajpai2025queue}), we deliberately avoided referencing existing these features by name, so that moderators who primarily used Old Reddit or the mobile app could still respond. Instead, we focused on intervention concepts that could plausibly serve these goals.

We first asked moderators about ways efficiency might be improved. These questions focused on what kinds of information or interface elements helped them act more quickly or with greater confidence, and included prompts about which report attributes shaped how long review typically took. Although not framed as interventions per se, these items provided context about what slows moderators down and what signals help streamline report review.

We then turned to interventions that address redundancy, asking for feedback on possible features that might reduce overlapping review. Two concrete ideas were presented: one was similar in spirit to the newer activity indicators feature on New Reddit \cite{reddit2024newmodqueue} that would display the real-time location of another moderator currently viewing a report in the queue (rather than only signaling after an action was taken as is the case for the activity indicator feature). The other was a more novel idea in which moderators would by default see a curated subset of the items queue (but could still access the full report set); the idea here was that if multiple mods were not shown the same reports at the same time, then collisions could be avoided, though we did not explain this reasoning to moderators so as to guage authentic responses for this idea without biasing with potential outcomes. For both of these proposals, moderators were given the option (and in some cases, required) to explain their answers in open-ended form boxes. 

Finally, to understand more generally how new features should be presented within the modqueue interface, we asked respondents to compare two modalities: inline visual cues (e.g., markers or signals embedded directly in the queue) and structured sorting or filtering options that reorder the list of reports. We selected these modalities because they are supported consistently across Old Reddit, New Reddit, and the mobile app, making the questions broadly applicable regardless of which version of Reddit participants used.

\subsection{Recruitment and Deployment}

We recruited moderators from a broad range of subreddits to capture diversity in experience, community type, and reliance on the modqueue. Following strategies used in prior survey research on Reddit moderation communities \cite{lambert24positive,bajpai2025queue}, 
we applied the following filtering procedures:
We began by identifying five public communities dedicated to moderation—r/ModSupport, r/modhelp, r/modnews, r/AskMods, and r/modclub—and collected usernames of Redditors who had posted in these forums between January 1 and April 1, 2025. For each user, we retrieved the list of subreddits they actively moderated. Subreddits were included if they were at least three months old, had at least 500 members and no less than two human moderators, and which averaged ten or more comments per post in the prior month. After filtering, we obtained a pool of over 1600 eligible subreddits. From this set, we randomly selected 1,400 to receive survey invitations (so that we could complete recruitment within our targeted timeframe), which were distributed via Reddit’s modmail system to each moderation team. Messages were sent periodically, with particular effort taken to avoid sending too many modmail messages in a short time frame so as to avoid Reddit API rate limits and Reddit's spam triggers. Survey deployment occurred over a three-week period, concluding at the end of April 2025.

To encourage participation, we offered an optional raffle in which one \$25 Amazon gift card was awarded for every 50 participants who opted in, a compensation scheme consistent with prior survey studies of Reddit moderators \cite{koshy2023alignment,lambert24positive,bajpai2025queue}.

\subsection{Analysis}

Because the majority of survey items were closed-ended, our analysis primarily relied on standard descriptive statistics (e.g., counts, proportions, means) to summarize patterns across responses. These results provided a baseline view of how moderators rated different objectives, interventions, and modalities. For the open-ended questions, the first author conducted a review and coding pass. The focus of this analysis was to surface illustrative rationales, contradictions, or noteworthy perspectives that enriched interpretation of quantitative findings. 

\section{Survey Findings}
\label{sec:findings}

We received 106 completed responses to our survey, collected through responses from 98 unique subreddits that were contacted during deployment. Respondents collectively oversaw 396 subreddits of different sizes and subjects. Respondents’ moderation experience ranged from under a year to two decades. Our respondent pool also captured a wide variety of Reddit version preferences.


\subsection{Moderators’ Perspectives on Objectives}\label{sec:findings:objs}

\begin{figure}[htbp]
\centering
\includegraphics[width=0.85\linewidth]{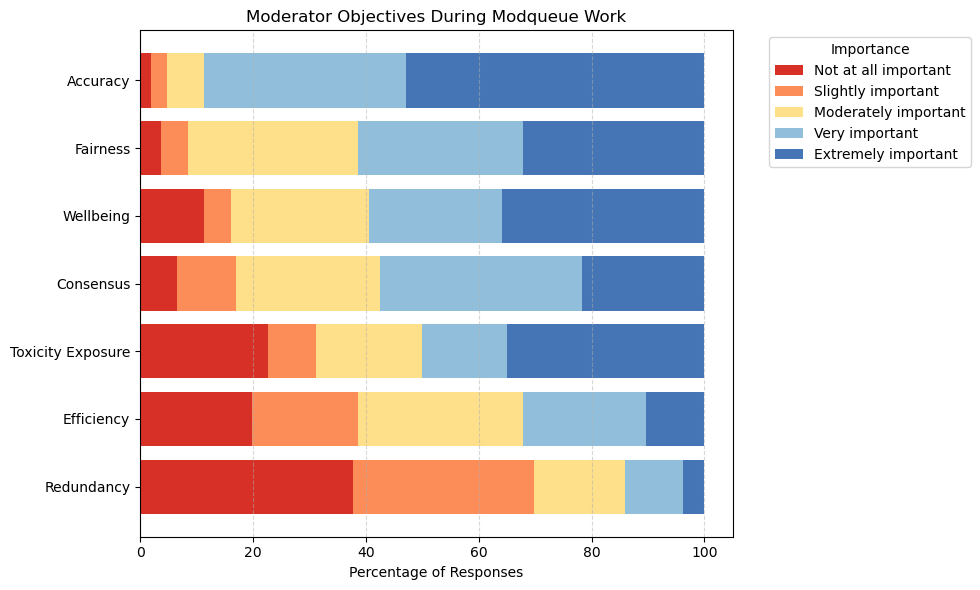}
\caption{
Moderator ratings of different report review objectives (5-point Likert scale). 
Accuracy received the highest average importance rating ($\mu=4.35$), followed by Fairness ($\mu=3.81$) and Wellbeing ($\mu=3.68$). 
Objectives such as Consensus, Toxicity Exposure, and Efficiency showed greater variation and lower mean ratings, and Redundancy had the lowest average importance rating overall ($\mu=2.10$). 
These patterns highlight differences in how moderators prioritize different goals during modqueue work.
}
\label{fig:mod-objs}
\end{figure}

Amongst the objectives that we directly asked about (shown in \Cref{tab:mod-obj-prompts}), Accuracy stood out as the most consistently valued objective, receiving the highest average rating, followed by Fairness and Wellbeing (\Cref{fig:mod-objs}). In contrast, both Efficiency and Redundancy received the lowest average ratings, despite Reddit’s recent emphasis on these objectives motivating new feature enhancements \cite{reddit2024newmodqueue}.

\subsubsection{Highest Rated Objectives: Accuracy, Fairness, and Wellbeing}

\paragraph{Accuracy.}
In open-ended responses, mods emphasized that accurate report decisions are essential not just for rule enforcement but also for maintaining legitimacy, building trust with users, and reducing downstream moderation work. Several noted that accuracy in moderation decisions is especially important in subreddits dealing with sensitive topics such as health, grief, or misinformation. For example, one respondent explained: 

\begin{quote}
\emph{I mod pet health and pet grief sub-reddits. Mistakes can let information through that can cost the life of an animal, mistakes can fail to let information through that could save the life of an animal. Knowing when to allow rule violations may save the life of an animal.}  (P015)
\end{quote}

Other respondents pointed out that lapses in accurate report review can lead to negative consequences for the overarching health and stability of the community; P020 noted that \emph{``removing content that does not violate policies creates mistrust.''}

\paragraph{Fairness.}
 For many, ensuring that the community's perception of fairness in report decisions was seen as a way to maintain community cohesion and reduce future moderation burden. For instance, P024 wrote \emph{``if there isnt a perception of fairness, the community will dissolve.''} Additionally, P110 provided a detailed account of how perceptions of unfairness have the potential to derail the discourse within subreddit entirely:

\begin{quote}
\emph{In my experience, perceived sense of unfair moderation catches fire quite easily, and ... sways the hot theme of the subreddit from the interesting subject to political matter of the community (P110)}
\end{quote}

Still, a few mods disagreed with fairness as an important objective in favor of more precise or objective metrics such as accuracy or consistency:

\begin{quote}
    \emph{``Fair'' is an arbitrary and capricious term, and therefore worthless. ``Ensure moderator decisions are made in accordance with and in adherence to the rules posted on the subreddit'' is a much better standard. It ensures compliance, uniformity in decision making and no favoritism. (P029)}
\end{quote}

\paragraph{Wellbeing.}
 In open-ended responses from mods that stressed the importance of wellbeing, mods referenced burnout, repeated exposure to disturbing content, and the emotional labor involved in decision-making. As P041 puts it: \emph{``Burnout is real, and this is a volunteer position, no one’s well-being should be compromised for any role, especially one they’re not being compensated for.''}

Others, however, pushed back against framing wellbeing as something that should be valued or worked towards. Several responses suggested that moderators are volunteering by choice, and that if someone cannot handle the demands of review work, they should opt out. For instance, P067 expressed: \emph{``Mods are volunteering. If they can't handle it they can step down or take a break.''} Another mod (P045) dismissed concerns about mod wellbeing during report review more blunty, stating \emph{``It's just words, bro. Sticks and stones can break my bones, but words don't hurt my feelings.''}


\subsubsection{More Contested Objectives: Toxicity Exposure, and Consensus}

\paragraph{Toxicity Exposure.}
Respondents who rated Toxicity Exposure as ``extremely important'' linked it directly to protecting mods' wellbeing, especially in communities where hateful, graphic, or otherwise harmful content is more likely to be reported. Some respondents described toxicity exposure as a serious risk that platforms and mod teams should actively work towards mitigating, especially for more vulnerable mods: 

\begin{quote}
 \emph{I want to look out for my fellow mods. I want to catch the use of the n word before a moderator of colour has to see it, I want to mute the people who might say awful things in modmail. Protecting my team is important to me. (P019)}
\end{quote}

Others, however, echoed the views expressed by those who opposed Wellbeing, emphasizing that moderation is voluntary and that emotional resilience is part of the role. 

\paragraph{Consensus.}
The importance of Consensus also drew mixed reactions. Some participants emphasized the value of team cohesion, consistency, and presenting a united front, especially when helping onboard new moderators or maintaining clarity in community enforcement. As P012 put it: \emph{``If we don't agree we will argue about stuff when we shouldn't.''}

Others, particularly those on solo teams or working with mostly inactive teammates, found the idea of consensus less relevant. A few noted that disagreement was inevitable, and that review work often happens asynchronously anyway, leaving little opportunity to solicit others' opinions until after decisions have been implemented. As P083 explained, \emph{``That agreement comes after the moderation decisions through internal communication.''} In such cases, consensus was viewed less as a live objective during modqueue use and more as something that would be handled retroactively, and only if needed.

\subsubsection{Lowest Rated Objectives: Efficiency and Redundancy}

\paragraph{Efficiency.}
Along with efficiency being one of the lowest ranked objectives, it drew the greatest variation of ratings across respondents. Some moderators, especially those working in high-volume communities, viewed quick triage as essential to keeping up with moderation demands and preventing burnout. They emphasized the need for smooth workflows, manageable queues, and features that could reduce friction. For example, P011 wrote \emph{``Nobody wants to spend hours in the modqueue. I have been there and it's extremely exhausting.''} 

In contrast, other mods that opposed the importance of efficiency felt that it would undermine more important objectives such as Accuracy or Fairness. Few mods noted that they moderated subreddits where volume was low enough that efficiency was not a concern. A few explicitly rejected the premise that mods should worry about review efficiency, suggesting that moderation should not be ``speedrun'' and that the time-consuming nature of moderation work is something volunteer moderators should come to accept if they take on the role: 
\begin{quote}
    \emph{It takes a long as it takes to do it correctly to the best of ability. Don't volunteer for a tedious and time consuming task and then complain it takes too long, rush through, and fail. Do the job, it takes as long as it takes. (P029)}
\end{quote}

\paragraph{Redundancy.}
Despite being ranked the lowest amongst respondents, a handful of respondents rated it as extremely important. These moderators emphasized that redundant effort not only wastes time but can also introduce inconsistency if overlapping decisions are made. For example, P092 indicated that it was \emph{``frustrating to waste time working the same queue item when we could have addressed another queue item during that time.''} P085 tied the issue more directly to consistency, explaining that such situations should be avoided ``for consistency and to limit multiple messages to the user,'' with the latter reason referencing how notifications or direct messages would be sent to users if multiple, contradicting decisions are made on their content.

The majority of respondents marked redundancy as not at all important. For many, particularly those moderating smaller subreddits, redundancy was simply not a common problem. Some explained that they were the only active moderator at a given time, or that moderators in their community worked during different ``shifts:'' 

\begin{quote}
    \emph{[Redundancy] does not matter. That is not a significant loss of efficiency since we are a rather small subreddit and oftentimes do not have multiple moderators working at once. (P014)}
\end{quote} 

Other respondents noted that the time wasted by occasional redundancy amounted to only seconds or minutes, hardly worth prioritizing. Some mods, like P019, noted that so long as mods were "being consistent, it shouldn't matter who addresses it," once again underscoring the connections to consistency.

\subsubsection{Other Objectives Surfaced from Open-Ended Response}


A common theme across responses (and that was repeatedly referenced in the open-ended feedback from those who had strong opinions regarding accuracy, fairness, consensus, and redundancy) was the importance of \textit{consistency}. Several moderators emphasized the need to apply rules uniformly, not only to uphold community expectations but also to foster trust within moderation teams. One mod described it as the foundation for helping newer moderators build confidence in decision-making and avoid interpersonal disputes over edge cases.

Closely related to consistency were values around \textit{community alignment} and \textit{transparency}. Some respondents highlighted the importance of making decisions that reflect subreddit norms, rather than simply enforcing rules in a vacuum. These mods also stressed the need to clearly communicate actions to users (whether through modmail responses, public logs, or well-written removal explanations). 
In a few cases, mods explicitly surfaced the goal of \textit{timeliness,} not necessarily to maximize speed or improve efficiency, but to ensure that reports are addressed before they escalate.

\subsection{Moderators' Insights Regarding Interventions}\label{sec:findings:inter}

\subsubsection{Signals that Increase Review Efficiency}\label{sec:findings:inter:time}
 
To gauge moderators’ insights on interventions or features targeted at improving review efficiency, we first asked about the time it typically takes them to review items in the modqueue. We include these responses since they may be of interest for future studies regarding improving modqueue efficiency. Respondents reported a wide range of review times. The majority (96\%) indicated that their average review time for a single queue item was under three minutes, with 54 selecting “less than one minute” and another 49 selecting ``1–3 minutes.'' Both the minimum and maximum times reported, however, varied significantly. The median shortest time reported was 10 seconds, with a few moderators reporting that they could resolve simple cases in under 5 seconds. On the other end of the spectrum, the median longest time was 10 minutes, though a handful of outliers reported review sessions that stretched to hours or even multiple days—the most extreme case involved a report that took over 10,000 minutes (roughly one week) to resolve. 

When asked which report attributes most influenced the time it takes to review a queue item, respondents most frequently selected the subject matter of the reported content (71\%), the need to gather additional context (71\%), and the projected complexity of the decision (64\%). Other common influences included the length of the reported content (50\%) and whether the report involved a repeat offender (45\%). In open-ended responses, moderators emphasized that ambiguous rules or unfamiliar users often required more deliberation, and that complex cases sometimes prompted team discussion, further extending review time. 


\subsubsection{Insights into Interventions to Prevent Collisions}\label{sec:findings:inter:aware}

\paragraph{Awareness Indicators} Mods expressed broad support for features that would help increase awareness of what other moderators are actively reviewing. When asked directly whether they would want to know if another moderator was currently reviewing a report, 78.2\% of respondents said yes. Similarly, a majority described such features as useful, with 61.8\% rating a live activity indicator as moderately, very, or extremely useful. 

 In open-ended responses, one mod mentioned the utility of being able to see whether a report had already been opened or assessed by someone else not just to reduce redundant review, but also to signal potentially complex cases:

\begin{quote}
    \emph{``Seen by X mods'' showing you that several mods have viewed and not actioned a report, indicating additional discussion or consensus needed. (P029, describing additional information should be presented in the modqueue)}
\end{quote}

\paragraph{Default Queue Filtering}

Respondents seem to push back on the idea of filtered queue defaults: When asked whether mods should be shown a filtered subset of reports as a default, while having access to the full set, responses were split: 63 disagreed (38 strongly) and 43 agreed. In contrast, in response to the statement ``moderators should, by default, see all reports from the subreddits they moderate,'' 94 respondents agreed, including 67 who indicated strong agreement; only 8 respondents disagreed. 

In open-ended follow-ups, many moderators pointed to the potential issues with of showing only a subset of reports by default. Some framed full visibility as critical for preserving context during decision-making, with P098 explaining, \emph{`I'd worry that if I wasn't able to see all of the reports in the queue, I would not understand the contextual reasons for the problems.''} As P018 noted, \emph{``We should see all reports, in order to ensure that they are being taken care of, and to determine any patterns.''} 

Others who assumed that Reddit would come up with the filtering questioned the platform's ability to do so reliably: P021 indicated that they \emph{``don’t trust Reddit to curate a list of reports for me.''} but that they would \emph{``trust a curation of [their] own making.''} 

Despite the majority of respondents being against the idea, some moderators were open to filtered views in specific contexts:

\begin{quote}
    \emph{It would be nice to be able to have trial mods only deal with lower-level reports, or indeed be able to say to someone ``you can just avoid dealing with racists in your volunteer job, let us handle that for you.'' (P019)}
\end{quote}

Others expressed that certain reports should be made only visible to veteran mods to ensure proper handling. For instance, P024 expressed \emph{``I like the idea of tiered moderator levels, where some mods only see some baseline, and other things can be left to the higher tier mods.''}

\subsubsection{New Visual Cues Are Preferred Over Sorting and Filtering Mechanisms}\label{sec:findings:inter:vc}

When asked about how new information to help improve the modqueue experience should be presented in the modqueue, a clear preference emerged: 74.5\% of respondents said they would rather have additional information surfaced through visual cues than through sorting or filtering options. Likert responses reinforced this trend. Only 18.8\% of participants rated new sorting or filtering options as ``very'' or ``extremely'' useful, while 39.6\% rated new visual cues in the same top categories.

Open-ended responses further clarified this preference. Mods often described the value of lightweight, inline signals that could reduce the need for extra clicks or cognitive load. For example, P011 indicated that \emph{``having different colors for reported content''} that highlighted particular report information would help them work faster; similarly P065 said \emph{``it would be useful to have a visual indicator of the length of a post or comment''}  when asked about what features or information they would like to have in the modqueue.

Despite the majority preference for visual cues, some open ended responses from mods referenced sorting or filtering in proposals for features to improve modqueue work:

\begin{quote}
    \emph{Moderators should be incentivized to use the modqueue, so the amount of time needed to review reports should be minimized in any way possible, such as by optimizing filters. (P014)}
\end{quote}

\subsubsection{Feature Suggestions Emphasize Modularity and Flexibility}

A recurring theme in participants’ responses---one that could already be inferred from recent survey findings~\cite{bajpai2025queue} 
and the diverse landscape of moderation preferences on Reddit---was the importance of giving moderators control over new features and the ability to create modular and customizable workflows. Rather than imposing platform-wide defaults, many suggested that features should be opt-in, adjustable, or individually configurable. For example, P056 explained that they would prefer the ability to \emph{``toggle for this to appear in the UI''}, since they did not believe the features they found useful \emph{``would be useful as a standard feature for everyone.''} Others noted that certain features might only be useful in specific circumstances. P018, for instance, remarked that new capabilities \emph{``could help during events or trends,''} suggesting that temporary or situational toggles could make features more broadly acceptable without disrupting typical workflows. Overall, respondents interested in new features stressed that modqueue features should accommodate different team structures, moderation styles, and subreddit needs.

\section{A Simulation-Based Approach to Auditing Modqueue Practices and Interventions}\label{sec:sim}

Our survey reveals an intricate picture of what it would mean to ``improve'' the modqueue. In addition to providing actionable design recommendations (provided in \Cref{sec:disc}), our findings point to several complexities and trade-offs that shape how moderators themselves understand and evaluate improvements to their workflows. To navigate these complexities, in this section we present a simulation-based approach to auditing modqueue practices and intervention proposals.

We use simulations to study moderation processes that are difficult to capture empirically, offering a way to explore underlying dynamics in greater depth through controlled experimentation and preempt the efficacy or pitfalls of current approaches and new interventions. Further motivation for pursuing simulations are discussed in \Cref{sec:disc:impr}.
    

\subsection{Agent-Based Model for Simulating Moderator Review on the Modqueue}\label{sec:sim:over}

We present a simulation-based audit tool for modqueue work, implemented as an agent-based model \cite{smaldino2023modeling}
in which moderators are represented as agents acting on a shared set of reports. Our focus is on one specific workflow or usage type uncovered by \cite{bajpai2025queue}: mods work through items in the modqueue continuously until every item has been reviewed and resolved, at which point the queue is empty and the review session ends. Prior work \cite{bajpai2025queue} also found several other ways in which mods use the modqueue, but we set those aside here in order to scope the simulation clearly.

To make the simulation tractable, we additionally adopt several simplifying assumptions (and reflect on our decision to do so in \Cref{sec:disc:lfw:simp}). We treat the modqueue as a static batch of reports, with no new items arriving during a run. All moderators see the same set of reports, though their ordering could differ to represent different default sortings depending on mods interface or tooling configurations. Once review begins, moderators’ orderings remain fixed: they do not re-sort or re-filter. Reports have a fixed review time, and all moderators work at the same pace. These assumptions provide a baseline for exploring moderator strategies and coordination dynamics, and can be relaxed or extended in future work. 

The main components of our agent-based model are as follows:

\paragraph{Reports \boldmath$\mathcal{R}$ and Moderators \boldmath$\mathcal{M}$}
We denote the report set as $\mathcal{R} = { r_1, r_2, \dots, r_N }$ and the set of moderators as $\mathcal{M} = { M_1, M_2, \dots, M_k }$. 

\paragraph{Modqueue Views  \boldmath$\Pi$ and Report Selection Strategies \boldmath$\mathcal{S}$a}
Each moderator has a \emph{modqueue view} which is the order in which reports are presented. In simple workflows (e.g., clearing top to bottom), this also reflects the order mods conduct review work in, but in some cases the two diverge \cite{bajpai2025queue}. To capture this, our model assigns each moderator both a modqueue view and a report selection strategy (RSS). We define the set of views as $\Pi = { \pi_1, \dots, \pi_k }$, where $\pi_i: [N] \to \mathcal{R}$ gives the report order shown to moderator $M_i$. The set of strategies is $\mathcal{S} = { S_1, \dots, S_k }$, where each $S_i$ is a probability distribution over positions in the queue that determines how items are selected. A strategy might be to always pick the first unreviewed item, favor earlier positions, or sample more evenly across the list. Changes to $\Pi$ or $\mathcal{S}$ can thus represent sorting/filtering features (\Cref{sec:findings:inter}) or alternative prioritization strategies used by moderators \cite{bajpai2025queue}.

\paragraph{Modqueue Traversal Process}
As is standard in agent-based models \cite{smaldino2023modeling}, the simulation unfolds in discrete time steps, which structure how agents (mods) make decisions, take actions, and update the state of the overall system; in our case, these steps capture how reports change status (from incomplete to completed, as we will discuss below) as moderators select, review, and resolve them. At each time step, any moderator who is idle uses their report selection strategy (RSS) to select a position within their modqueue view. The mod then attempts to begin reviewing the corresponding report, as long as it is still incomplete at the time step they select it. In the next section, we will see how our simulations can be modified to emulate the effects of introducing interventions, in particular awareness indicators (\Cref{sec:findings:inter}) or changes to default sorting, that can, with some probability, prompt mods to select different positions in the queue. Once a moderator begins reviewing a report in our simulation, they remain occupied for a fixed number of time steps (equal to the report's length) unless the report is completed by someone else in the meantime.

\paragraph{Report and Mod Attributes}
Each report $r_i \in \mathcal{R}$ is assigned a length $len(r_i)$, indicating how long it takes to review (assumed to be the same across all moderators in this initial formulation). In \Cref{sec:disc:lfw} we discuss how additional attributes can be incorporated to model different concerns or complexities. 

\paragraph{Report Completion}
A report $r_i$ in our simulated modqueue is considered completed once any moderator reviews it for $len(r_i)$ time steps. We will (for now) assume mods stop working on a report as soon as some mod has completed it, unless stated otherwise. In Reddit’s actual modqueue, whether mods actually would stop their review might depend on whether mods are aware of the fact that an action has already been taken for it. 

\paragraph{Simulation Process}
At the beginning of each simulated run, all reports (that are pre-specified at the start of the run) will be marked as incomplete and all moderators will be idle and ready to begin reviewing. Time proceeds in discrete steps. At each step, idle mods use their report selection strategy to probabilistically choose a position in their modqueue view and attempt to begin reviewing the corresponding incomplete report. Moderators who are already busy continue reviewing their current report until the required time has elapsed and the report is marked as complete. Once a report is completed, any other moderators working on it are released and return to idle. Pseudocode for this process is provided in \Cref{alg:abm}, which can be found in \Cref{app:pseu:core}, and sample python code for implementing these simulations can be found in the first author's Github. 


As noted earlier, the above is not intended to capture every factor shaping modqueue work, but to serve as a starter model to show how one can experiment with the modqueue review process in a systematic way. In the next section, we illustrate how this simulation can be configured and used for experimentation to gain more insights into review work via the modqueue.

\section{Experimentation Using Simulations: Proof of Concept}\label{sec:exp}


In this section, we show how using simulations alongside empirical findings or anecdotal accounts (such as those from \Cref{sec:findings} and \cite{bajpai2025queue}), can provide insights into current modqueue practices and the likely effects of interventions on objectives moderators could care about. More broadly, this illustrates how simulations can serve as a tool for disentangling the unseen connections or tradeoffs that can accompany efforts to ``improve'' the modqueue.

\subsection{Sample Experiment 1: Exploring the Relationship between Redundancy, Efficiency, and Wellbeing}\label{sec:exp:one}

Prior work \cite{bajpai2025queue} and our survey findings (\Cref{sec:findings}) highlight discrepancies in how moderators view collisions and their impact on efficiency. Some respondents described collisions as largely inconsequential, suggesting they have little effect on the overall time required to process reports. Others, especially those on larger moderation teams, emphasized the frustrations of repeatedly encountering collisions, linking them to wasted effort and diminished efficiency. To probe these relationships systematically, we configured a set of simulation experiments to examine how collisions could affect efficiency and how increasing the size of a moderation team may influence these dynamics. 

We additionally consider the potential implications of collisions for moderator workload as it pertains to Well-being, as some open-ended responses that emphasized the importance of efficiency and moderator wellbeing indicated that diminished efficiency (which may be caused by the prevalence of collisions) could contribute to fatigue and, in the long run, lead to moderators experiencing burnout (\Cref{sec:findings:objs}).

\subsubsection{Operationalizing Objectives as Metrics}\label{sec:exp:one:objs}

As a first step in configuring our simulation experiments, we translate the objectives we intend to explore into concrete metrics that can be captured and tracked within simulated trials. Below, we present four such metrics that we will use.

\begin{itemize}
    \item \textbf{Modqueue Completion Time:} The total number of discrete time steps required for all reports in the modqueue to be completed. 
    \item \textbf{Collision Count:} Total number instances where a moderator begins reviewing a report that is already being reviewed by another. 
    \item \textbf{Mod Work vs. Redundant Time}: \emph{Work time} is the amount of they spend during the simulated trial working on reports; if mods work continuously without interuptions, this will simply be an individual mods' completion time. \emph{Redundant time} for each mod can capture the amount of time units they spend working on reports that were already being reviewed. 
    \item \textbf{Report Seen vs. Completed Counts:} \emph{Reports completed} by each mod are the number of reports they are responsible for completing during a simulated trial (i.e. the reports that they review and complete before anyone else does). \emph{Reports seen} by each mods includes those they may begin to review but do not complete. 
\end{itemize}

These metrics are not meant to be exhaustive or definitive ways of operationalizing the objectives, but should be understood as partial heuristics that capture certain dimensions of them. 

\subsubsection{Experiment Setup}\label{sec:exp:one:set}

 For the sake of simplicity, in every trial we instantiate the modqueue with 100 incomplete reports. We also assume that all reports are of uniform length (in this case, all reports will have length 5 to allow for sufficient time for collisions to potentially take place). We know from our survey findings in \Cref{sec:findings:inter:time} that this is not a realistic assumption (since report complexity and review time vary considerably based on our respondents' experiences) but we adopt it here to streamline our experiments. We discuss these assumptions further in \Cref{sec:disc:lfw:simp}. 

With respect to configuring the set of mods for each trial, we decided to vary the total number of active moderators between 2 and 10 to explore how team size may shape the three objectives we are probing. All moderators in a given trial are shown the same ordering of reports, but to introduce variability in their behavior we assign them a probabilistic report-selection strategy: at each step they select the first incomplete report in their queue with probability 0.6 and the second incomplete report with probability 0.4. This reflects prior findings that moderators tend to work sequentially through the queue \cite{bajpai2025queue}, while leaving room for small deviations. As we will discuss in \Cref{sec:disc:lfw}, in practice moderators employ a range of traversal strategies, and these can be implemented and experimented with in more extensive simulation studies.

As is standard in simulation-based experimentation, we run multiple (100) trials for each configuration (i.e. each combination of parameters we vary; here, we only have nine configurations, since the only parameter that is being varied are the number of moderators actively working in the queue). Repeated trials allow us to account for the stochastic variation introduced by the probabilistic selection strategy and to examine the distribution of realized outcomes, rather than relying on a single run. As we will see in the experiment results, we do not see a wide range of variation in our trial outcomes within each configuration; this is because we only have one parameter that introduces variation. Once again, fully fleshed out simulation studies should involve far more exhaustive experimentation \cite{smaldino2023modeling}.  

\subsubsection{Results} \label{sec:exp:one:res}

\begin{figure}[ht]
\centering
    \includegraphics[width=\textwidth]{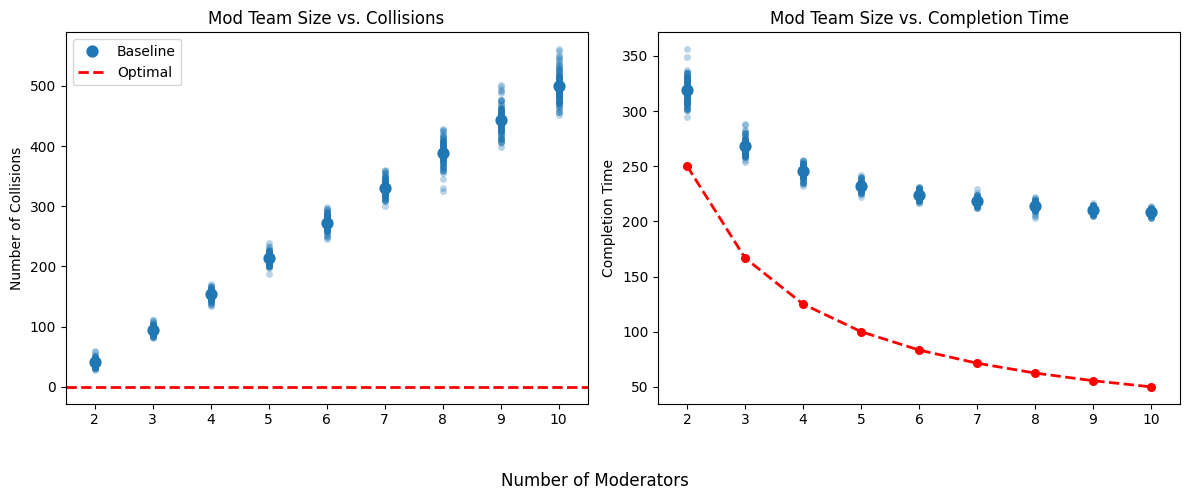}
    \caption{(\textit{left}) The number of collisions increases as the number of active mods increase. (\textit{right}) The optimal completion time of the modqueue decreases as the number of active mods increase. Notice that completion time of our baseline does not improve as dramatically with larger moderation teams as is the case for the optimal completion time values.}
     \label{fig:time-and-colls}
\end{figure}

\begin{figure}[tbh]
\centering
    \includegraphics[width=\textwidth]{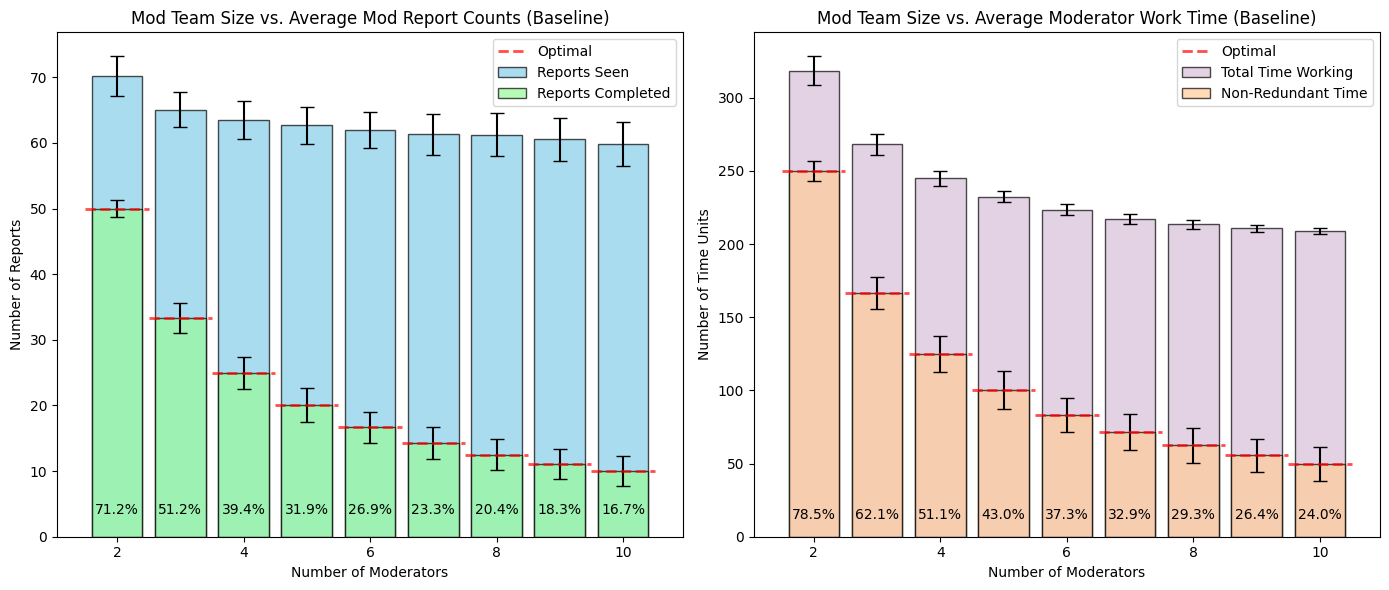}
 \caption{(\emph{left}) Average number of reports completed by an individual mod during a trial compared to the average reports seen; percentages describe the average percentage of reports seen that end up completed. (\emph{right}) Average amount of time an individual mod spent working compared to the amount of time that actually contributes towards realized work (i.e. finished reports), or non-redundant time; percentage values indicate the percentage of total working time that is spent non-redundantly. Note that mods are spending time on and seeing more reports than those they complete, indicating that much of their actual effort is spent doing work that could be seen as ``unnecessary'' as it does not contribute to completing reports.}
     \label{fig:workload-defaults}
\end{figure}

The results of our experiment trials with respect to the metrics we defined above can be seen in \Cref{fig:time-and-colls} and \Cref{fig:workload-defaults}. In both figures, a red-lined is used to denote what the ``optimal'' outcomes would be with respect to each metric. Here, we are using ``optimal'' to mean what the lowest possible value could potentially be for each trial configuration. For instance, the optimal completion time in trials with 2 mods will be $(100 \times 5) \div 2 = 250$.  These will establish a benchmark for estimating room for improvement. 

There is a substantial gap between the performance of our simulated trials and the optimal values across the board. On average, it takes moderation teams $2.70$ times longer than the best-possible completion time to complete reports. The largest performance gap occurs when there are $10$ mods working concurrently; the optimal completion time in this scenario is $50$, whereas our experiments indicate have an average completion time of $208.3$ which is over $4$ times larger. Furthermore, in the optimal scenario, we see an $80\%$ improvement in completion time when increasing the moderation team size from $2$ to $10$, but only see a $34.6\%$ improvement in the actual results of our experiment. For collisions, we see an average of $57.8$ more collisions when the moderation team increases by one. 

Based on how we set our experiment up, the overall work time metric coincides with the modqueue completion metric. 
As such, we will instead focus on the report counts metrics. While the average number of reports completed matches the optimal scenario, the variance of these quantities can be used to discern whether large discrepancies occur across individual mods within a trial. On average, we see a variance of $9.21$ and $5.26$ for the total number of reports seen and reports completed, respectively, indicating instances in which some mods see over $10$ or over $5$ more reports than others. We observe that while the total number of reports completed decreases as the number of mods increase, the number of reports mods pick up to review plateaus. We find the largest gap between the number of reports seen vs. completed when there are 10 mods, where on average each mod sees $59.8$ reports, they only review $10$ (or $16\%$) of them to completion. In other words, a mod may be opening and viewing $60$ reports, but will only contribute in completing $10$ such reports.

\subsection{Sample Experiment 2: Probing the Efficacy of Potential Interventions}\label{sec:exp:two}

Our survey findings surfaced moderators’ preferences regarding interventions to improve modqueue workflows. To build on these findings, we use our simulation to probe the potential efficacy of two potential interventions that could potentially support mods' objectives.

The first intervention we evaluate is an awareness indicator (\Cref{sec:findings:inter:aware}) which could be implemented within modqueue interfaces can be implemented as a lightweight visual cue integrated into the modqueue interface that signals when another moderator is already viewing or working on a report. The second is the use of alternate sorting mechanisms. Despite visual cues being a more popular modality for modqueue interventions amongst our survey respondents (\Cref{sec:findings:inter:vc}), some mods did see value in surfacing information via sorting features. Additionally, prior work \cite{bajpai2025queue} found that some mods already alter the order in which they review reports as a way to prioritize certain objectives. 

For the sake of simplicity, we will continue to focus on efficiency and redundancy as the primary objectives we are exploring, but discuss how simulations may be extended to explore other objectives in \Cref{sec:disc:lfw}). 

\subsubsection{Simulating Interventions}\label{sec:exp:two:inter}

Because our simulation does not replicate the modqueue interface itself, but rather the underlying process of moderators working through reports, interventions must be encoded in terms of how they may alter the simulated system behavior. Below, we describe how we represent each intervention within our simulation and specify the parameters and configurations used for experimentation. 

\begin{itemize}
\item \textbf{Awareness Indicators (Visual Cues).} We introduce a parameter $\rho$ to model how moderators react when their chosen report is already under review. If $\rho = 0$, they always continue with the initial choice. If $\rho > 0$, they proceed with probability $1-\rho$ and otherwise skip to the next item. This serves as a proxy for the effectiveness of visual cues in steering moderators away from redundant work.

\item \textbf{Alternative Sorting Configurations.} To model interventions that change report ordering, we vary moderators’ \emph{views} of the modqueue (the order in which reports appear). We note that we could have alternatively changed mods' selection strategies as a way to simulate this intervention as well. We compare three configurations: \textbf{Uniform}, where all moderators see the same order; \textbf{Reverse}, where half see the list forward and half in reverse; and \textbf{Random}, where each moderator sees a random permutation.
\end{itemize}

We use the same trial configuration from \Cref{sec:exp:one:set} (i.e., 100 reports of uniform length 5, and probabilistic selection strategies between the first and second incomplete report), but will vary interventions as well as the number of moderators actively reviewing. We also continue to evaluate outcomes using two of the metrics defined earlier in \Cref{sec:exp:one:objs}: modqueue completion time (for efficiency) and collision count (for redundancy). 

\subsubsection{Results}\label{sec:exp:two:res}

\begin{figure}[ht]
\centering
\includegraphics[width=\textwidth]{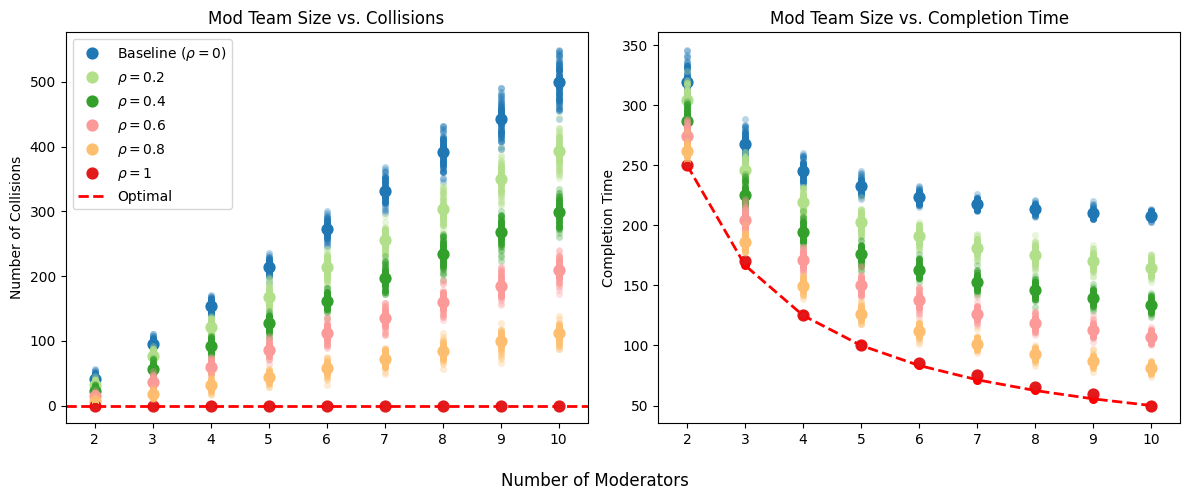}
\caption{(\textit{left}) Increasing collision awareness reduces the number of collisions. (\textit{right}) Completion time also improves with higher $\rho$, reaching optimal values when $\rho = 1$.}
\label{fig:colls-time-aware}
\end{figure}

\begin{figure}[ht]
\centering
\includegraphics[width=\textwidth]{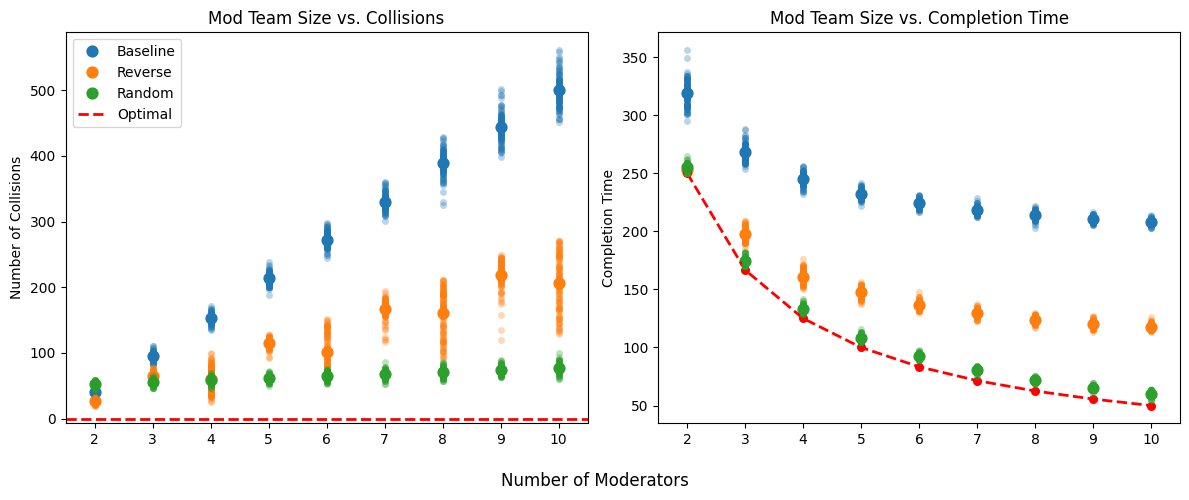}
\caption{(\textit{left}) Reverse and Random traversal strategies reduce the number of collisions. (\textit{right}) These same strategies also lead to faster completion times, with Random achieving performance close to the theoretical optimum.}
\label{fig:colls-time-views}
\end{figure}

Results from our simulation trials are shown in \Cref{fig:colls-time-aware,fig:colls-time-views}. In awareness indicator experiments, we see that when $\rho = 1$ trials achieve optimal performance on both metrics. While such perfect awareness is unlikely in practice, even modest increases in $\rho$ produce meaningful gains. For example, raising $\rho$ from 0 to 0.2 reduces collisions by 20\% and improves completion time by 10.8\% on average across mod team sizes.

For alternate sorting configurations, both interventions outperform the baseline (\textit{Uniform}) on efficiency and collision metrics. On average, the \textit{Reverse} configuration reduces collisions by 50.3\% and improves completion time by 36.2\%. The \textit{Random} configuration produces even larger improvements, reducing collisions by 61.3\% and improving completion time by 53.6\%. These gains are especially notable given that moderators in all simulations continue to use the same selection strategy. In the \textit{Random} case, average completion time comes within 10.7\% of the optimal benchmark.

\section{Discussion}\label{sec:disc}

\subsection{Design Implications and Recommendations for Improving Modqueue Review}

\subsubsection{Customizable Contextual Signals to the Modqueue Can Support Diverse and Evolving Review Work}

Our survey findings underscore the importance of contextual signals within the modqueue interface. Respondents indicated that their decision-making often relies not only on explicit metadata (e.g., report type or time) but also on more implicit forms of context that may not be immediately obvious to platform designers or researchers. For example, respondents emphasized the value of scanning the full queue (not simply to track overall moderation activity of communities, as highlighted in \cite{bajpai2025queue}), but because patterns across reports directly shape how they interpret individual items and decide on appropriate actions.

These insights suggest that modqueue improvements should focus on surfacing contextual information in lightweight, easily interpretable ways. Inline visual cues could be particularly attractive to respondents, because they could provide high-level signals without requiring moderators to have to search for or synthesize specific contextual information. While newer contextual and user panel features provide users with large amounts of information, open-ended responses in our survey indicated that they may not need \emph{all} of the information that is shown. Instead, a more curated and digestible set of information could be provided. Examples of potential signals that are not yet provided by the modqueue (as of writing this paper), but could be useful for mods include the number of times a report had been viewed but not acted upon (to help signal potential complexity), the reported post's length, or statistics that provide an overview of user's behavior within the subreddit (e.g. the percentage of their contributions that have been reported or the number of flaired or highly upvoted content within a certain timeframe). 

Because both individual mods and subreddits have their own unique needs and norms, these signals should be configurable, i.e., mods should be able to pick and choose which types of signals are presented to them within the modqueue interface. Taking this idea a step further to support mods that may moderate multiple subreddits, platform designers could allow for these configurations to vary depending on the subreddit from which a certain modqueue item comes from. More concretely, consider a moderator (Maude) who helps run two very different subreddits, r/A and r/B. Because each community has distinct rules and norms, Maude may rely on different kinds of contextual signals to guide her decisions. In r/A, moderation often hinges on interpreting nuanced discussions and community tone, so Maude might benefit from cues that show surrounding conversation context and lightweight metadata about the users involved. In r/B, by contrast, rule enforcement may be more straightforward. Here, Maude could rely primarily on simpler signals such as the length of the report, or an automated toxicity indicator that flags excessive profanity, without needing the additional overhead of reviewing surrounding conversation.

This variation is not only community-specific, but also temporal, shifting as the conditions or norms within a subreddit evolve. Suppose r/B is suddenly featured on r/popular~\cite{chan2024understanding} and experiences a surge of new traffic and reports. In this situation, Maude might want different contextual signals than usual, such indicators of how many contributions a user (who has submitted a report) has made in the subreddit, to help distinguish newcomers unfamiliar with the rules from established members. User-specific signals of this sort could also help Maude detect potential report abuse, which prior research has shown to be a persistent challenge in many communities \cite{gilbert2023towards,are2025flagging}.

\subsubsection{Non-Technical Ways to Improve Modqueue Work Can Lead to Improved Outcomes}

Because prior studies \cite{bajpai2025queue} have shown that many moderators are content to traverse the queue in a simple top-to-bottom order, even simple variations in default item orderings can have potentially significant effects on moderation outcomes. Our simulation experiments in \Cref{sec:exp:two} suggest that varying the default order of reports across moderators is one way to distribute attention more effectively without dramatically altering individual workflows.

At the same time, these findings point to the potential of non-technical approaches. Lightweight \emph{coordination strategies}, such as assigning some moderators to start at the top of the queue and others at the bottom or leaving certain kinds of reports to more veteran moderators (or mods who have specialized expertise, which is already standard practice in communities like r/AskHistorians \cite{gilbert20historians}) could provide many of the same benefits without requiring platform-level changes. Such practices could be especially useful for mod teams that have more fragmented tooling preferences (which may diminish the efficacy or availability of features such as activity indicators \cite{modqueue}) but that still wish to improve their review work by limiting redundancy or inaccurate review decisions. 

\subsubsection{Reconciling Design Intent vs. Community Use}

A recurring theme across our findings is the gap between what tools might be designed for and how they are actually used in practice. Features are often motivated by particular goals (such as improving efficiency or reducing redundancy \cite{reddit2024newmodqueue}) but the diversity of moderation practices means that their real value may emerge in entirely different ways. As a result, tools that initially seem unnecessary or even counterproductive may still find meaningful use in specific contexts.

For example, we proposed default filtered modqueue views as a way to reduce the potential for collisions to occur by distributing moderator attention across different subsets items within the queue. While this design could indeed help in this regard, respondents raised strong concerns that such filtering would obscure valuable context and limit their ability to detect broader patterns across reports. At first glance, this seems like a feature that simply should not be pursued. Yet survey responses also surfaced more targeted use cases, and specifically those that would not be as upended by the lacking of broader queue context: filtered views could help onboard new moderators by presenting them with simpler reports, or protect vulnerable team members by filtering out especially harmful content from their modqueue views. 

This contrast underscores the importance of designing tools that can accommodate multiple modes of use. Rather than assuming a one-to-one alignment between design intent and moderator needs, platform designers should expect that features will be repurposed, reinterpreted, or even resisted depending on the community context. Furthermore, these variations should be explored more thoroughly: platform designers should make more of a concerted effort to understand how certain tools may effect moderation processes across different categories of subreddits, rather than leaving communities to be responsible for identifying such potential.  Building in modularity, transparency, and opportunities, and, more importantly, making clear what potential uses might look like, is therefore integral to ensure that tools not only support their original objectives but also adapt to the varied practices and challenges that define volunteer moderation.

\subsubsection{Interventions Should Be Explicitly Motivated to Encourage Adoption}

Moderators may not always be able to accurately anticipate which tools will genuinely help their communities or improve their moderation practices. As a result, new features often face resistance not only because moderators are accustomed to existing workflows, but also because proposed interventions may not appear immediately useful. This hesitation or resistence does not necessarily mean the tools lack value; rather, it highlights a gap between what a feature can accomplish and how its potential is communicated to moderators.

One way to bridge this gap is by providing moderators with more tangible evidence of a feature’s effectiveness. As we will discuss further in \Cref{sec:disc:lfw}, it would be valuable to study whether showing moderators concrete outcomes (such as data demonstrating reduced collisions, faster queue clearance, or fewer redundant reviews) would make them more willing to try and subsequently experiment with new tools. Doing so could increase adoption, but more importantly, realize new ways to adapt them for their specific needs. This process of adaptation can generate new practices or ideas for improvement, and reveal unanticipated benefits that extend beyond the original design goals.

For platform designers and tool creators, this suggests that the rollout of new features should not only emphasize availability and intended goals, but also explicitly demonstrate value. Making the benefits of interventions visible and verifiable can help overcome initial skepticism, encourage experimentation, and ultimately support the evolution of moderation practices over time.

\subsection{Implications for Researchers: Agent-Based Simulations as Transparent Proxies for Complex Processes}\label{sec:disc:impr}
Our survey findings align with prior literature on content moderation and 
emphasize how varied and context-dependent even seemingly “universal” moderation processes can be. Practices differ across communities, between moderators within the same team, and can evolve over time as community norms shift. This fluidity makes it unrealistic to expect a single, definitive account of what ``reviewing reports'' looks like in practice, which subsequently makes it harder to find ways to improve or support this type of work. 

While this variation could, in principle, be uncovered or further investigated through detailed empirical studies of the modqueue, doing so in practice could prove to be especially difficult. Controlled experiments on live subreddits might raise ethical concerns around intervening in real communities, and would be complicated by the fact that moderators rely on different Reddit interface versions and external tools that vastly change what features are available to mods (as is the case with Reddit's modqueue) \cite{bajpai2025queue}. Tracking real usage of the modqueue (including report information and moderation decisions) could be seen as invasive to moderators, and even if moderators consented to such tracking, isolating the effects of a single intervention would be difficult given the number of contextual factors shaping moderation work. Moreover, moderators willing to participate might represent only a narrow slice of moderation philosophies and practices, limiting insights into the broader applicability of the resulting findings.

Simulation offers a way to navigate these constraints. By constructing simplified modqueue instances, researchers can systematically vary parameters such as queue composition, traversal strategies, and potential interface cue effects, or experiment with variations regarding mod team size and composition to probe how different conditions or cases may shape outcomes. This allows us to surface potential issues or tradeoffs that might otherwise go unnoticed, and to translate moderators’ expressed values into testable hypotheses. Our own experiments illustrate how even highly abstract simulations can reveal structural bottlenecks (such as collisions) and highlight downstream effects on metrics that speak to objectives such as efficiency or wellbeing.

At the same time, simulations should not be viewed as replacements for empirical study, but as complementary tools. In this paper, survey findings revealed specific discrepancies (such as moderators downplaying the effects of collisions) that simulations were able to interrogate further. As we will discuss in \Cref{sec:disc:lfw}, simulation findings can in turn be probed or validated using targeted empirical studies. These more targeted empirical insights, whether gained from interviews, trace analyses, or co-design workshops, can ground simulation parameters in lived practice, while simulation models can in turn extend those insights to explore hypothetical interventions at scale.

In addition, using agent-based models \cite{smaldino2023modeling} provide a more transparent and explainable way of simulating underlying workflow processes. In contrast with simulations powered by generative AI, our approach foregrounds the mechanics of review and makes it easier to dissect and recreate outcomes, allowing for more concrete ways of validating or auditing experiments. An additional benefit of this is that results could be more saliently and transparently explained to moderators or community owners, allowing them to engage with simulated results more meaningfully. 

Together, these implications suggest that progress in studying the mechanics of moderation workflows could benefit from hybrid approaches: empirical methods to capture the richness of real practices, and simulation methods to systematically vary conditions and probe interventions in a low-risk environment.  Our experiments provide a cursory view of how this can be done, 
but there is a need for
richer modeling, which we will discuss in \Cref{sec:disc:lfw}. 

\subsection{Limitations \& Future Work}
\label{sec:disc:lfw}

\subsubsection{Deeper Investigations into Survey Findings}

While our survey allowed us to capture a wide range of perspectives from moderators across different communities, it still provides only a surface-level view of why certain objectives or interventions are valued over others. For example, we cannot determine whether moderators’ emphasis on particular objectives is shaped by the types of subreddits they moderate, the number of subreddits they are responsible for, or other contextual factors such as team size or community norms. Additionally, it fails to capture situations where mods may value certain objectives or features when moderating a specific subreddit, but value different objectives when conducting review for other subreddits. Future work can conduct in-depth interviews to examine these underlying reasons more directly and understand how values and objectives are formed and prioritized in moderation work.

Our survey also did not attempt to address or engage with mods that did not regularly use the modqueue, since mods that began taking our survey but indicated that they spent no time using the modqueue were simply shown the end of the survey. However, investigating what might encourage such mods to begin using the modqueue could be valuable for understanding how to improve the modqueue and support moderator review work in general. Once again, future studies via in-depth interviews or participatory-design could center around gaining a deeper understanding of what features or changes might make the modqueue more useful for mods, or establishing how certain mod review workflows outside the modqueue can be improved. 

In short, while our survey surfaced important differences how mods value objectives and view potential improvements, more targeted studies are needed to uncover deeper motivations behind these differences and to ensure that future designs for the modqueue account for the full range of moderator needs.

\subsubsection{Transitioning From Proof-of-Concept to Richer Simulation Experiments}
\label{sec:disc:lfw:simp}

As noted throughout \Cref{sec:sim,sec:exp}, our simulation framework and sample experiments were deliberately simplified in order to maintain clarity and explainability in order to clearly communicate the value of simulation-based methodology as part of the broader goals of this paper. While our simplifying assumptions made the experiments and result easier to explain and present, they do not (yet) faithfully reflect the dynamic nature of real report review, where new items can arrive at any time, vary widely in complexity, and are handled by moderators with different levels of expertise and availability. Similarly, the discrete time units that governed our simulations were not mapped to real measures of time or productivity, making it easy to argue that the quantitative outcomes of our experiments are not grounded in reality.

We highlight these as limitations to once again emphasize how our current experiments should be viewed as proof-of-concept demonstrations rather than as realistic predictions or fully-fleshed out simulation studies. Future work should refine these experiments by incorporating greater variation in report attribute parameters (e.g. length and complexity) and how they impact review time or moderator selection strategies. Likewise, future simulations should more carefully calibrate time units to empirical measures, making outcomes more interpretable in practice.

A promising path for improving our sample experiment design would be to focus on a single subreddit where the moderation team relies heavily on the modqueue. Through a targeted empirical study (using either survey, interview, or trace data if available) researchers could estimate modeling parameters such as the distribution of report types, average review completion times, and moderator attributes that mods indicate influence their decision-making. These community-specific insights could then be used to instantiate a more realistic model, which in turn could be varied systematically to probe how interventions or alternative practices might play out in that particular context. This narrower scoping would allow the results from simulation experiments to better motivate actionable insights, rather than simply illustrate broad phenomena.

Such refinements should not stop at varied parameterization. Simulation results can and should be further validated. This can be done by engaging moderators directly (e.g., showing them simulation outputs and eliciting whether the dynamics align with their lived experiences), or by deploying lightweight prototypes of interventions and observing their effects in practice. These forms of validation would help ensure that simulation-based findings remain both credible to moderators and useful to platform designers.


While our simulation in \Cref{sec:sim} is deliberately simple, its modular design makes it straightforward to extend. New attributes and process variants can be layered in to capture additional aspects and objectives of modqueue work. Future work could extend the model by adding parameters and sub-processes for moderator expertise and accuracy, where report complexity and domain knowledge dictate how likely reviews are to be correct. Adding new modeling components could also capture how exposure to toxic content degrades moderators' review performance, while arrival-time parameters could be used to guage delays or urgency in report review when different traversal strategies are used.

\section{Conclusion}

In this paper, we examined what it means to “improve” Reddit’s modqueue by focusing on the objectives moderators bring to review and their preferences regarding potential interventions. Our survey of 106 moderators surfaced a wide variety of perspectives on both objectives and interventions. While moderators consistently valued accuracy in report review, they expressed contrasting views on other objectives; others noted how certain objectives are intertwined, which points to potential trade-offs making it difficult to prioritize one without undermining another. Respondents favored lightweight, flexible features that could complement existing workflows (such as visual cues), and were more opposed to ideas that would takeaway contextual signals. Given the diversity of our survey findings and the challenges of measuring modqueue outcomes directly, we turn to simulations to investigate how practices, objectives, and interventions interact within modqueue workflows. 

Overall, our findings suggest that supporting moderator review requires designs that are flexible, modular, and responsive to diverse objectives. Platform designers should build tools that can be adapted or configured to account for variation in community context and moderator values, but must also recognize how improvements made with respect to one objective may introduce costs to others. For researchers, our work shows how simulation can complement empirical methods by enabling systematic experimentation and deeper analysis of underlying modqueue and review dynamics.

\appendix

\section{Simulation Pseudocode} \label{app:pseu}

\subsection{Core Simulation Dynamics} \label{app:pseu:core}

\begin{algorithm}[tbh!]\label{alg:abm}
\caption{\textsc{Overview of Agent-Based Simulation for RR}}
\DontPrintSemicolon
  \SetKwFunction{Define}{Initialize}
  \SetKwInOut{Input}{Input}\SetKwInOut{Output}{Output}
 \SetKwBlock{Begin}{begin}{}
 Initialize Report Set $\mathcal{R} := \{ r_1, ..., r_N \}$ where each $r_i$ has report length $len(r_i)$ and is marked as $incomplete$
 
 Initialize Mod Team $\mathcal{M} := \{ M_1,...,M_k\}$ where each $M_i$ has modqueue view $\pi_i$ and selection strategy $S_i$.

 Initialize time counter $T := 0$
  \BlankLine

  \While{there are incomplete reports in $R$}{
  \ForEach{$M_i \in \mathcal{M}$}{
    \lIf{$M_i$ is Idle}{
      \textsc{DoWhenIdle($M_i$)}  \tcp{i.e.  \Cref{alg:idle}}
    }
    \lElseIf{$M_i$ is Busy with a report $r$}{
    \textsc{DoWhenBusy($M_i$)} \tcp{i.e. \Cref{alg:busy}}
      }
    }
    $T = T + 1$\;
  }

\Return{$T$ as the amount of time it takes for mods in $M$ to complete the reports of $R$}

\end{algorithm}

\begin{algorithm}[tbh!]\label{alg:idle}
\caption{\textsc{DoWhenIdle}}
\DontPrintSemicolon
\SetKwInOut{Input}{Input}

\Input{Mod $M_i$ that is marked as \emph{idle}}

\eIf{no incomplete reports in $R$}{
    $M_i$ stays \emph{idle}\;
}{
    $r \leftarrow \textsc{SelectReport}(M_i)$ \tcp{Function that implements a selection strategy as described in \Cref{sec:sim:over}}
    $M_i$ begins working on $r$\;
    $M_i$ becomes \emph{busy} for $len(r)$ time steps\;
}

\end{algorithm}

\begin{algorithm}[tbh!]\label{alg:busy}  
\caption{\textsc{DoWhenBusy}}
\DontPrintSemicolon
\SetKwInOut{Input}{Input}

\Input{Mod $M_i$ that is working on report $r$ for $t$ time steps}

$M_i$ works on $r$ for current time step\;
$t = t + 1$\;

 \If{$t == len(r)$}{ \tcp{i.e. $M_i$ has worked on $r$ for the required amount of time}
        $r$ is marked as completed\;
        $M_i$ becomes idle\;
        \ForEach{$M_j \in M$ that was busy with $r$}{ \tcp{Release all other busy mods that were working on $r$}
          $M_j$ becomes idle\;
        }
        
    }
      
\end{algorithm}


\bibliographystyle{ACM-Reference-Format}
\bibliography{references}

\newpage


\end{document}